\documentclass[10pt,a4paper,prl,twocolumn]{revtex4}
\usepackage{amsmath}
\usepackage{amsfonts}
\usepackage{amssymb}
\usepackage{graphicx}
\begin{document}

\title{Comment on ``Granular Entropy: Explicit Calculations for Planar Assemblies''}
\author{Massimo Pica Ciamarra}\email[]{picaciamarra@na.infn.it}
\affiliation{Dipartimento di Scienze Fisiche, Universit\'a di
Napoli `Federico II', CNR-Coherentia, INFN, 80126 Napoli,
Italia.}
\homepage{http://smcs.na.infn.it}
\begin{abstract}
\end{abstract}

\maketitle

In the statistical mechanics approach introduced by S.F. Edwards to describe granular media in their
mechanically stable states, the granular entropy is defined as the logarithm of the number
of mechanically stable configurations $N$ grains can be arranged in a volume $V$.
In order to calculate this entropy, it is necessary to introduce a volume function $W$
expressing the total volume as a function of the position and coordination of the $N$ grains. 
Recently Blumenfeld and Edwards~\cite{Blumenfeld} have proposed such a volume function in two dimensions.
This volume function is constructed by first partitioning the system into voids defined from the grain contacts, and then by partitioning the voids into quadrilaters defined from the grains centers and the position of the contact points. 
The volume of the system is equal to the sum of the volumes of the voids, i.e. to the sum of the areas of the quadrilaterals.
Here I show  that the proposed partitioning of the voids into quadrilaterals is not exact, and therefore that the introduced volume function is not a volume function as claimed.

\begin{figure}[t!!!!!!]
\begin{center}
\includegraphics*[scale = 1.4]{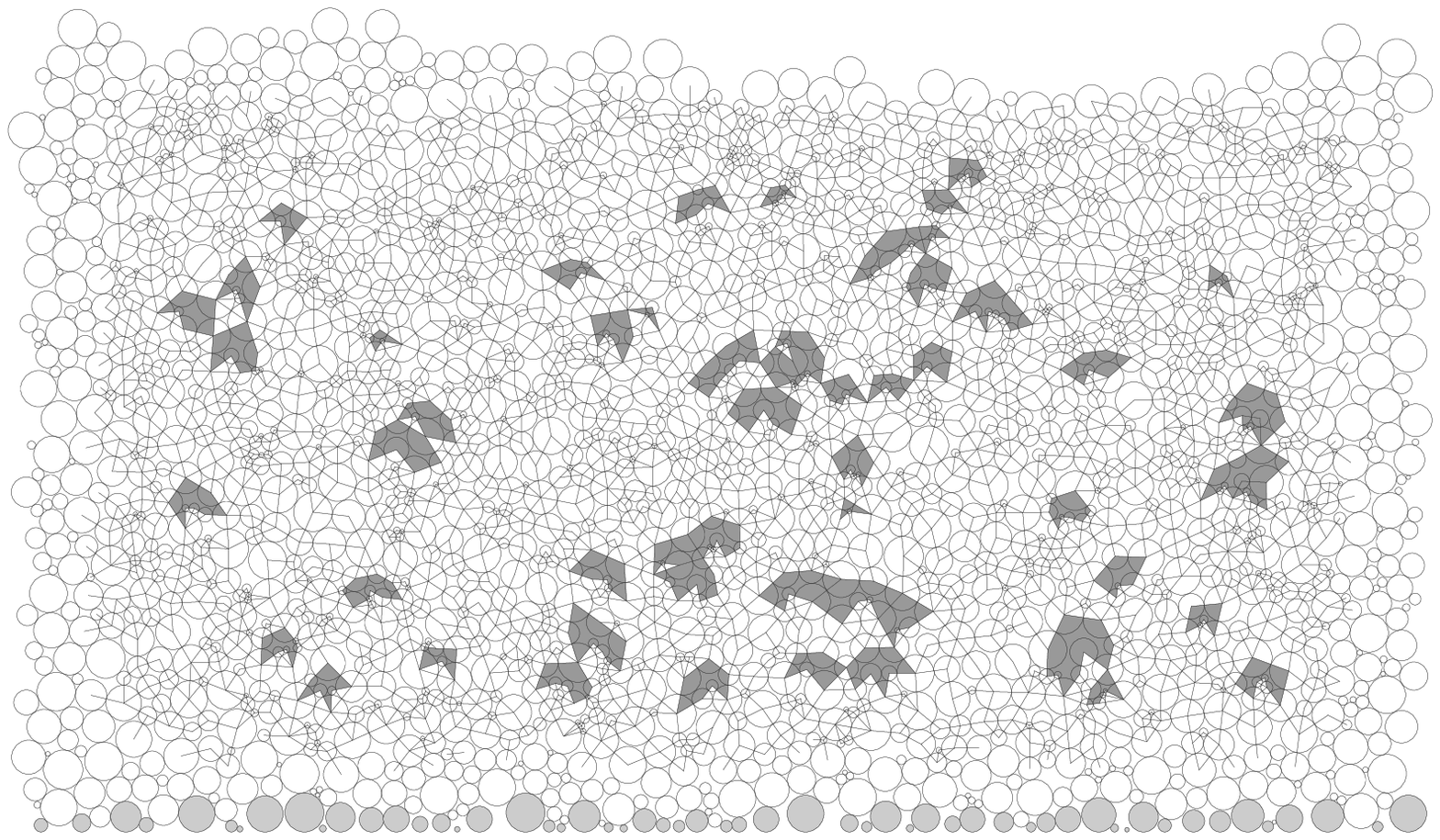}
\caption{\label{fig1} The central part of a mechanically stable (under gravity) configuration of a polydisperse system is tessellated in voids according to the procedure of Ref.~\cite{Blumenfeld}. The shaded non-convex voids are not well tessellated by the quadrilaters defined in Ref.~\cite{Blumenfeld}.}
\end{center}
\end{figure}

The partition of a void into quadrilaterals works when the void is convex, as that shown in Fig. 1 of Ref.~\cite{Blumenfeld}. Some non-convex voids, which has been overlooked in Ref.~\cite{Blumenfeld}, are instead not well partitioned.
In order to show that the presence of these non-convex voids is relevant, I have investigated the voids of mechanically stable (under gravity) configurations of 2D polydisperse granular systems, generated by simulating a sedimentation process via standard soft-core Molecular Dynamics simulations~\cite{Silbert}.
Fig.~\ref{fig1} shows (the central region) of one of the investigated configurations, together with the partition of the system into voids. In this configuration about $20\%$ of the voids are non convex, and $5\%$ of the voids (the shaded ones) are not well tesselleted by the quadrilaterals introduced in Ref.~\cite{Blumenfeld}.

Two parameters are of interest to evaluate the effect of non-convex voids on the analysis performed by Blumenfed and Edwards. One is the fraction $f$ of voids which are not well tesselleted, the other is the relative error in the estimation of the volume, $\eta = (W-V)/V$, where $V$ is the volume occupied by the grains, and $W$ its estimation.
Figure 2a and 2b shows the dependence of $f$ and $\eta$ on the polydispersity of the system~\footnote{The diameter of the grains is uniformly in $[d_{min},d_{max}]$, and  $p = (d_{max}-d_{min})/(d_{max}+d_{min})$.}, for $\mu = 0.2,0.4$ and $0.6$. 
The magnitude of $\eta$ is understood by considering as an example a monodisperse system of volume $V$ and area fraction $\phi$, with total volume $V =nv_s/\phi$,  $n$ number of particles and $v_s$ area of a particle.
In this case $\phi_{rlp} \simeq 0.84 < \phi < \phi_{crystal} \simeq 0.90$, and a very crude extimation of the total volume as  $W = nv_s/\phi_m$,  with $\phi_m = (\phi_{rlp} + \phi_{crystal})/2$,  gives a relative error $\eta < 4\%$.

These results indicates that $f$ and $\eta$ increase both with $p$ and with $\mu$, and that for spherical grains the volume function introduced by Blumenfeld and Edwards works only at first crude approximation, even in the absence of polydispersity. The presence of non-convex voids will be certainly more relevant when non-spherical grains are considered. 

\begin{figure}[t]
\begin{center}
\includegraphics*[scale = 0.38]{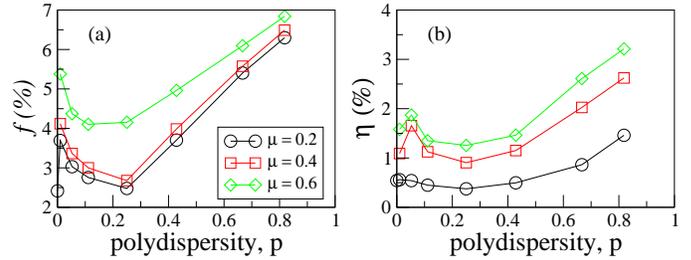}
\caption{\label{fig2} (a) The fraction of voids $f$ not well partitioned by the quadrilaterals introduced in Ref.~\cite{Blumenfeld}; (b) The relative error made when measuring the volume with the volume function introduced in Ref.~\cite{Blumenfeld}.}
\end{center}
\end{figure}

I thank T.Aste, A. Coniglio and M. Nicodemi for helpfull discussions.


\begin{thebibliography}{40}
\bibitem{Blumenfeld} R. Blumenfeld and S.F. Edwards, Phys. Rev. Lett. {\bf 90}, 114303 (2003).
\bibitem{Silbert}L. E. Silbert, D. Ertas, G.S. Grest and T. C. Halsey and D.
Levine, Phys. Rev. E {\bf 65},, 051307 (2002).
\bibitem{Blumenfeld2} R. Blumenfeld and S.F. Edwards, Eur. Phys. J. E, {\bf 19}, 23 (2006).

\end{thebibliography}
\end{document}